\documentclass[fleqn,usenatbib]{mnras}

\usepackage{newtxtext,newtxmath,comment}
\usepackage{bm}
\usepackage[T1]{fontenc}
\usepackage{ae,aecompl}

\usepackage{graphicx}	% Including figure files
\usepackage{amsmath}	% Advanced maths commands

\usepackage{amssymb}	% Extra maths symbols
\usepackage[normalem]{ulem}

\title{A Cost-Effective Search for Extraterrestrial Probes in the Solar System}

\author[Villarroel et al.]{
\and
Beatriz Villarroel,$^{1}$\thanks{E-mail: beatriz.villarroel@su.se} Wesley A. Watters,$^{2}$ Alina Streblyanska,$^{3}$ Enrique Solano,$^{4}$
Stefan Geier,$^{3,5}$
\and
Lars Mattsson$^{1}$
\\
$^{1}$Nordita, KTH Royal Institute of Technology and Stockholm University, Hannes Alfvéns väg 12, SE-106 91 Stockholm, Sweden\\
$^{2}$Whitin Observatory, Dept. Physics \& Astronomy, Wellesley College, 106 Central Street Wellesley, MA 02481, USA\\
$^{3}$ Instituto de Astrof\'{\i}sica de Canarias, V\'{\i}a   L\'actea, 38205 La Laguna, Tenerife, Spain\\
$^{4}$Departamento de Astrof\'isica, Centro de Astrobiolog\'ia (CSIC/INTA), P.O. Box 78, E-28691 Villanueva de la Ca\~{n}ada, Spain; esm@cab.inta-csic.es\\
$^{5}$ GRANTECAN: Cuesta de San Jos\'{e} s/n, 38712 Bre{\~n}a Baja, La Palma, Spain\\
}

\date{Accepted XXX. Received YYY; in original form ZZZ}

\pubyear{2025}

\begin{document}
\label{firstpage}
\pagerange{\pageref{firstpage}--\pageref{lastpage}}
\maketitle

\begin{abstract}
For centuries, astronomers have discussed the possibility of inhabited worlds — from Herschel’s 18th-century observations suggesting Mars may host life, to the systematic search for technosignatures that began in the 1960s using radio telescopes. Searching for artifacts in the solar system has received relatively little formal scientific interest and has faced significant technical and social challenges. Automated surveys and new observational techniques developed over the past decade now enable astronomers to survey parts of the sky for anomalous objects.
We briefly describe four methods for detecting extraterrestrial artifacts and probes within the Solar System and then focus on demonstrating one of these. The first makes use of pre-Sputnik images to search for flashes from glinting objects.  The second method makes use of space-borne telescopes to search for artificial objects. A third approach involves examining the reflectance spectra of objects in Earth orbit, in search of the characteristic reddening that may imply long-term exposure of metallic surfaces to space weathering. We focus here on a fourth approach, which involves using Earth's shadow as a filter when searching for optically luminous objects in near-Earth space.  We demonstrate a proof-of-concept of this method by conducting two searches for transients in images acquired by the Zwicky Transient Facility (ZTF), which has generated many repeated 30-second exposures of the same fields.  
In this way, we identified previously uncatalogued events at short angular separations from the center of the shadow, motivating more extensive searches using this technique. We conclude that the Earth's shadow presents a new and exciting search domain for near-Earth SETI.
\end{abstract}

\begin{keywords}
extraterrestrial intelligence -- transients -- surveys -- minor planets, asteroids, general
\end{keywords}

%%%%%%%%%%%%%%%%%%%%%%%%%%%%%%%%%%%%%%%%%%%%%%%%%%

%%%%%%%%%%%%%%%%% BODY OF PAPER %%%%%%%%%%%%%%%%%%

\section{Introduction}

Searches for Extraterrestrial Intelligence (SETI) using optical and radio telescopes have been going on for sixty years. This work has demonstrated remarkable technical advances alongside the development of creative search strategies and frameworks for discovery and rigorous evidential standards. Ever since astronomers understood that our own radio transmissions could be received by extraterrestrial astronomers \citep{CocconiMorrison1959}, radio observations have been running episodically, such as at the Green Bank telescope and Allen Telescope Array in California (United States). Despite this progress, no confirmed signal of extraterrestrial (ET) origin has been detected. Several research groups have made major contributions to radio searches: the SETI institute \citep[e.g.,][]{Tarter2001} have been engaged for decades. The Five-hundred-meter Aperture Spherical radio Telescope (FAST) in China has recently joined the search \citep{Li2020}. In 2015, the Breakthrough Listen program launched their effort to perform the most systematic and comprehensive radio search to date, observing 1 million stars, the galactic plane, and over 100 galaxies for radio signals. Their work has led to firm upper limits on radio emitters. For example, \cite{Enriquez2017} conducted searches for artificial signals in the 1.1--1.9 GHz range, operating at powers higher than $\sim10^{13}$ W, concluding that less than $<0.1$\% of all systems within 50 pc have radio transmitters this powerful. It should be noted that these limits apply to continuously transmitting and omnidirectional or Earth-directed transmitters.   \cite{Garrett2022} identified 143,000 galaxies that apparently also show no artificial radio signals, placing limits on the luminous end of the SETI luminosity function. Subsequent work by \cite{Price2020} searched for radio transmitters in the 1.10--3.45 GHz range, finding no transmitters stronger than $\sim10^{12}$ W. Again, these limits apply under the assumption of continuous and omnidirectional or Earth-directed transmission.  Other notable efforts include SETI Italia \citep{Montebugnoli2001} and searches conducted using the Russian RATAN-600 telescope \citep{Bursov2016}. Sixty years of radio searches have so far not yielded a candidate signal that satisfies agreed criteria for a detection of interest.

The lack of a clear discovery from radio searches has motivated searches for optical lasers, as they provide more privacy and higher bit rate than radio communication. Lasers used by extraterrestrial civilizations could have long-lived (continuous over minutes or hours or even longer) beams or emit short pulses, where the latter are much more energy efficient. These may also rotate or stay fixed with respect to the observer. One method to detect such a laser is to search for repeated, short pulses \citep[PANOSETI;][]{Wright2018,Maire2020}. An all-sky all-the-time search has been conducted by LaserSETI for the past 10 years\footnote{http://www.laserseti.net} \citep{Gillum2023}, where a network of small cameras survey the sky from different positions in search of a laser flash. Another tested method relies on examining the spectra of stars.  A powerful laser source that transits its host star is expected to leave a monochromatic emission line in stellar spectra \citep{Tellis2017}. For example, a 60 Megawatt laser with a 10 meter aperture laser beam at a distance of $\sim$30 pc, is detectable in the spectra of a host star above the continuum level---even when the laser pulse is as short as a second \citep[e.g.,][]{Marcy2022a}. Only limited parts of the entire parameter space have been sampled so far, and no candidate laser source has been identified up to now \citep{Stone2005,Howard2007,Reines2002,Marcy2021,Marcy2022a,Marcy2022b,Marcy2022c}.

In contrast, there have been few searches for extraterrestrial artifacts and probes in our Solar System. But this possibility was well understood since the early 1960s after interstellar spaceflight became realistic, raising the idea that ET civilizations might intentionally send probes (or unintentionally send other artifacts) to our immediate neighborhood \citep{Bracewell1973}. Carl Sagan suggested that an advanced extraterrestrial civilisation might send relativistic probes to our Solar System \citep{Sagan1963} and that the Earth might have been visited as many as $\sim10^{4}$ times. Indeed, while relativistic probes are extremely energy-demanding, it is not unimaginable that other civilisations have attempted sending probes similar to Voyager or Pioneer to other star systems, or even a simpler ``message-in-a-bottle'' style of artifact. In 2011, Voyager left the Solar System, and even with its non-relativistic, humble velocity, it is estimated to reach the distance of the closest star in about 77,000 years. These types of probes are far cheaper and more energy efficient for transmitting information to another civilisation \citep{Rose2004}, although they are arguably far more difficult to discover \citep{haqq2012likelihood}. In California, the 100 million dollar Breakthrough Starshot project was investigating new propulsion technologies to launch a probe with relativistic speed to the closest star (see e.g. \cite{Lubin2016}). Had they succeeded, the probe could reach the closest star in as little as twenty years. Sadly, the project was canceled.

An ET civilization may configure interstellar probes to land on the surfaces of planets \citep{Carlotto1990,Arkhipov1996,Davies2013}, or asteroids \citep{Papagiannis1978,Benford2019}, or else to park in a stable orbit around planets like Earth for millions of years \citep{Villarroel2022a}. Some probes may be intact, while others could have been around for hundreds of thousands of years and disintegrated over time. It has been speculated that thousands, millions, or even more probes might be spread throughout the galaxy, as part of a network of robotic spacecraft \citep{Schwartz1961,Freitas1980,Freitas1981,Zuckerman1985,Hippke2018,Hippke2020,Gertz2022}, some of which may communicate using lasers. In the early 1980s, searches for ET probes nearby the Earth were conducted with optical telescopes \citep{FreitasValdes1980,Valdes1983}, without yielding positive results. One study has been carried out using photographic plate material from the 1950s \citep{Villarroel2022c}, yielding two statistically significant preliminary candidates. The search for nearby probes and artifacts turned out to be the least-traveled path in SETI research, due to the heavy requirements of such searches, often based on astronomically expensive space missions.

There is an evident need for new, rigorous, and verifiable observations in the search for ET artefacts \citep{Shostak2020}. The recent efforts within academic institutions to investigate Unidentified Aerial Phenomena (UAP), although not necessarily related to ET technology, have brought the search closer to home (in Earth's atmosphere) through efforts like the \textit{Galileo} project \citep{Galileo,Watters2023, Domine2024}, UAPx \citep{Szydagis2025initial}, and IFEX (Interdisziplin\"aren Forschungszentrum f\"ur Extraterrestrik; \cite{Kayal2022,Kayal2023}).  

The search for ET probes exclusively \textit{outside} the Earth at geosynchronous orbits and beyond is motivated by a desire to avoid detecting national security assets and minimize false positives.  As mentioned above, the possibility that ET probes or artifacts are present in the Solar System is based on reasonable extrapolation from conservative estimates of spacefaring capabilities of advanced civilizations \citep{Sagan1963,Knuth2024}, and indeed from the examples of our own Voyager and Pioneer probes. Moreover, a handful of curious astronomical observations are at least somewhat suggestive that we should search closer to home, such as observation of the asteroid VG 1991 \citep{Steel1995}, the elongated object `Oumuamua \citep{Bialy2018}, and the finding of multiple transients in small regions of the sky \citep{Villarroel2021,Villarroel2022c,Solano2024}.  The most impressive case to date is arguably that of three bright ``stars'' appearing and vanishing within 50 minutes \citep{Solano2024}. The goal of finding ET artifacts and probes can be achieved through dedicated searches for specular reflections from the surfaces of artificial objects \citep{Lacki2019,Villarroel2022b} as well as intrinsic optical emissions. 

In this paper, we present search methods that permit us to filter out human-made satellites and other background noise, together with preliminary, proof-of-concept searches using telescopic images acquired in the past 8 years, using the Earth's shadow as a filter. 

\section{Filtering out the noise}

The greatest challenge when searching for ET probes and artifacts in interplanetary space using modern instruments is that our skies are polluted with thousands of satellites and millions of reflective pieces of space debris, even as more and more satellites are being launched into orbit. This contamination is an especially challenging problem for anyone wishing to search for non-anthropogenic objects. Below we present several solutions to separate the many glints and emissions caused by human-made objects from potential ET artifacts and potential ET probes and artifacts, and then we describe a demonstration of one of these. 

\subsection{Going back in time: using pre-Sputnik images}

A promising avenue is to examine digitized pre-Sputnik astronomical images captured by Palomar, Carte du Ciel, Harvard, and Lick observatories. These images, many of which date back to the early 1950s and earlier, can be analyzed for both anomalous events in the sky that occurred prior to the proliferation of satellites and space debris, and to search for typical signs of artificial satellites and space debris \citep{Villarroel2022a}. Discoveries made through this process might be usefully compared to the records of anomalous celestial observations made by astronomers before 1850, including those by Cassini and Messier as documented by \cite{vallee+aubeck2010}, adding to their interest and value.

Previously, \cite{Villarroel2022a} discussed a straightforward search method for specular reflections from artificial objects in geosynchronous orbits: multiple transients that are aligned as the object moves and rotates. The paper proposed that any alignments with more than four transients are particularly noteworthy and warrant further investigation to distinguish them from plate defects with atypical shapes (i.e., not point-like or extended due to movement). Although a few preliminary candidates have been identified \citep{Villarroel2022c}, the origin of these transients remains unknown. The events might be caused by an entirely different and unanticipated phenomenon than what the given paper tested (solar reflections from artificial objects in geosynchronous orbits), e.g. atmospheric or exoatmospheric, in which multiple stationary light sources appear and disappear from the field. A recent discovery of a bright triple transient event in Palomar plates on the 19th of July 1952 is particularly puzzling and convincing \citep{Solano2024}.

The Vanishing \& Appearing Sources during a Century of Observations (VASCO) project is currently planning a new generation of its citizen science project, which will search for glinting satellites instead of ``vanishing stars'' \citep{Villarroel2022b}.

\subsection{Observations with Space-Borne Telescopes}

An additional uncontaminated search approach involves examining a search volume that resides beyond the orbits of satellites and therefore beyond geosynchronous orbit. This approach can be achieved by using existing space telescopes, such as Kepler, or the Transiting Exoplanet Survey Satellite (TESS), and analyzing their existing datasets in search of anomalous objects.

TESS, for example, uses four wide-field cameras all pointed at the same field, making it an ideal tool for searching for rare events involving multiple or ``simultaneous'' transients, as reported in studies by \cite{Villarroel2021, Villarroel2022b} and \cite{Solano2024}. However, it is worth noting that events fainter than approximately 12 magnitudes may not be detectable by the telescope, which utilizes small mirrors.

While this experiment may not be able to detect any ET technological objects in orbit around the Earth, it does permit the search for objects farther out in the solar system.

\subsection{Color-coded Clues of Ancient Space Debris}

Those who wish to examine Earth orbits in modern-day data for dead artifacts without intrinsic emission will face a significant challenge: the view is spoiled by millions of human objects. Other projects, such as the Galileo project, tackle this challenge by searching for signs of interstellar meteors similar to 'Oumuamua in image data -- and even by exploring the ocean depths for potential artifacts following fireball events with potential hyperbolic atmospheric entry velocities \citep{Loeb+etal:2024}.

 An object that has resided in Earth's orbit for a long period of time will have been exposed to dust particles, cosmic rays, and micrometeorites, leading to a reduction in its reflectivity and causing a reddening of the object \citep[e.g.,][]{Hapke2001,pearce2020examining}. We can leverage this phenomenon to obtain the reflection spectra of the top 5\% most reddened artificial objects in Earth's orbit without a known ``owner'', meaning objects that are not listed by the USSTRATCOM. Of course, military satellites will not be in USSTRATCOM, which comprise a significant fraction. Nevertheless, solar reflectance spectra from an object will contain both the solar spectrum and a pattern of absorption features that are characteristic of the surface materials \citep{vasile2024space}. Therefore, spectra from the most reddened population, showing unusual absorption or emission lines that suggest an unknown material, could indicate strong candidates for ET artifacts.

\subsection{Using Earth's Shadow as a Filter}

The Earth's shadow is an ideal uncontaminated search space. Every night, the Earth casts a shadow cone where direct sunlight cannot reflect from satellites or space debris (although there is still a possibility of reflection from the much fainter Moonlight). The size of this shadow varies throughout the year, but the base has an average radius of around $8-9$ degrees for objects at geosynchronous orbits (GSO; approximately 35,700 km) \citep{Nir2024_git}. With typical velocities of artificial objects at GSO, which move at approximately 15 arcsec per second, a brief flash caused by intrinsic emission lasting only $0.2-0.5$ seconds can produce a point source, with its magnitude depending on the strength of its intrinsic emission.

The Earth's shadow is much larger at Low Earth Orbit (LEO), and objects emitting flashes move so quickly that they are likely to create long streaks as they rotate in most long-exposure imagery, rather than short flashes. Also, objects with continuous, longer-lasting emission further out than LEO might produce long streaks. It is therefore interesting to search for elongated streaks that occur within the shadow of the Earth and yet are well outside the atmosphere, in this way excluding aircraft and meteors. Such streaks are unlikely to be human satellites, which typically do not carry intrinsic optical light sources. The rare exceptions are largely confined to LEO and include (i) lasers used for communication (usually NIR) or laser ranging and atmospheric LIDAR; (ii) LEDs for optical self-identification; (iii) short-range LEDs used for optical tracking and formation flying; and (iv) spacecraft propulsion (e.g., rocket burns). At the distance of the Moon, an orbiting object will move at angular velocities of roughly $\sim0.5$ arcsec sec$^{-1}$. At a few times the distance of the Moon, we might search instead for streaks inside the shadow (which narrows down further as one moves away from Earth) indicating angular velocities $<0.1$ arcsec sec$^{-1}$. Determining the distance to streaking objects will require triangulation, as envisaged for the ExoProbe experiment\footnote{https://thedebrief.org/a-new-era-of-optical-seti-the-search-for-artificial-objects-of-non-human-origin/} \citep{ExoProbe2023}.

\begin{figure}
\includegraphics[width=20pc]{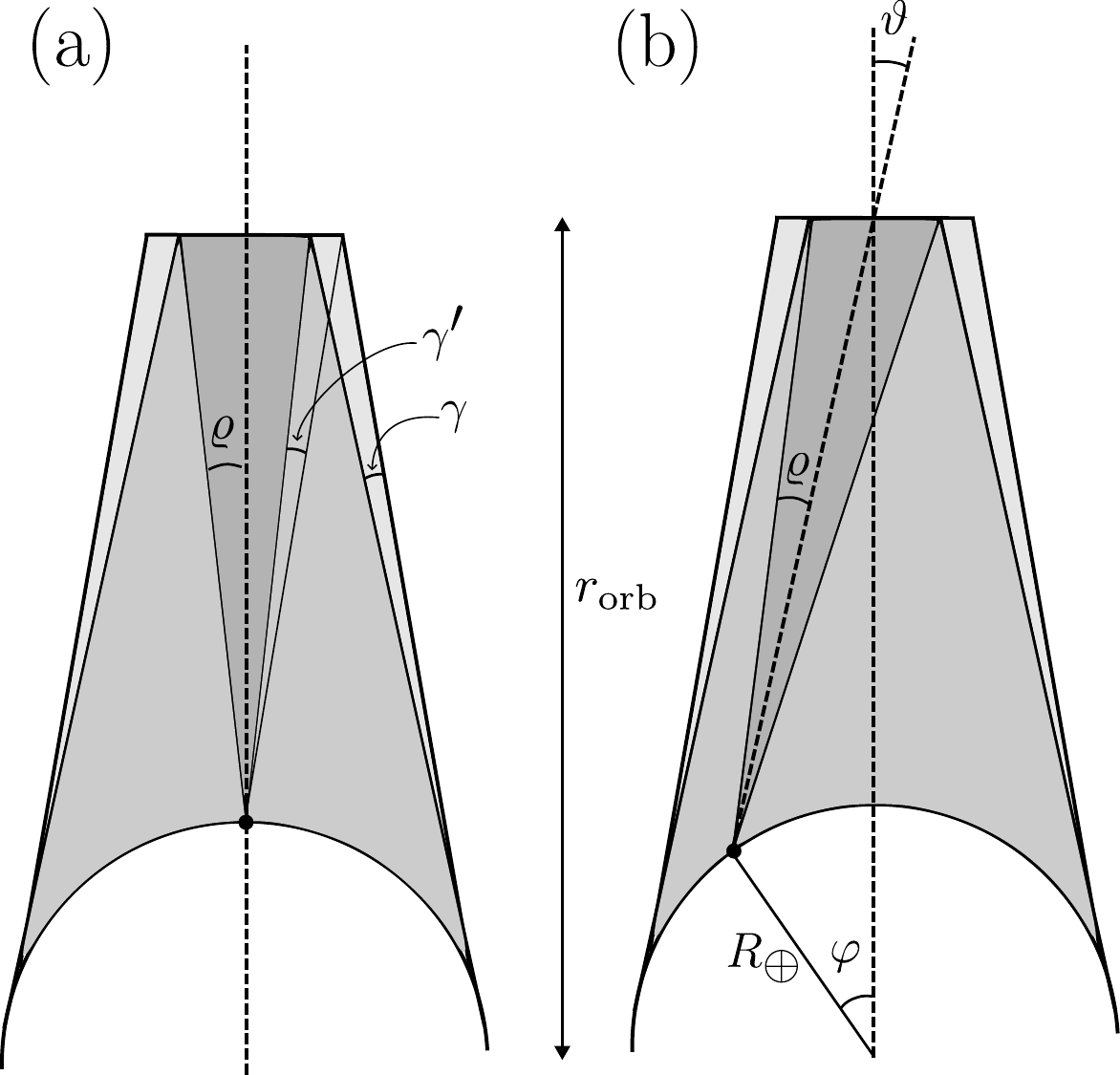}

\caption{Earth's umbra out to an orbit radius $r_{\rm orb}$ (not to scale).  The Sun
  is in the direction of the bottom of the page.  The lightest shading marks the ``refraction fringe'' into which sunlight is refracted by the atmosphere (angular size $\gamma$).  Intermediate shading is used to mark the ``dark full shadow'' (DFS).  The darkest shading highlights the shadow volume visible to an observer on the Earth's surface (this shading implies nothing about darkness in this volume, which is the same as the DFS). DFS radius is given by $\varrho$; parallax of the shadow center with respect to the antisolar point is $\vartheta$; angular separation of the observer from the geocenter-antisolar axis is $\varphi$. (a) Shows the shadow geometry for an observer on the Earth's surface beneath the antisolar point (the ``subantisolar point'' or SAP), and (b) shows the shadow geometry for an observer at an arbitrary position on Earth's surface, with angular separation $\varphi$ from the SAP.}\label{fig_shadow}

\end{figure}

The shadow of the Earth consists of the penumbra and umbra or ``full shadow''; the latter is depicted in Figure \ref{fig_shadow}.  The full shadow consists of a dark core (the ``dark full shadow'' or DFS) and an outer fringe of light refracted by the Earth's atmosphere, which can produce satellite glints and which measures $\gamma \sim 2^{\circ}$.  Assuming $r_{\rm orb} = 35,786$ km for geosynchronous orbit, the largest fractional change in observer position with respect to this altitude is $\Delta R/R \sim R_{\oplus}/r_{\rm orb} \approx 0.18$, which is hence also the largest fractional change in apparent angular size of the refraction fringe ($\gamma^{\prime}$) and DFS radius ($\varrho$) as a function of position on the Earth's night side.  At the geocenter, the total umbra radius is $\varrho + \gamma = \sim 8$ to $9^{\circ}$; from Earth's surface below the antisolar point (Figure \ref{fig_shadow}a), $\varrho + \gamma \approx 10^{\circ}$.  A ``safe'' shadow radius (for avoiding satellite glints) is therefore below $ 8^{\circ} - 2^{\circ} = 6^{\circ}$.  Alternatively, this could be calculated as a function of position $\varphi$ to maximize the search radius.  In this initial study, in most cases we applied a cutoff of $\rho < 6^{\circ} < \varrho$ for the radial angular separation ($\rho$) from the shadow center when we filtered detections.  

At distances far enough from Earth, such as those greater than or equal to 0.01 astronomical units (au), any detectable flashes will occur from beyond the Earth's shadow cone. Since very few human-made objects are located at these distances and their positions are well documented, there is a negligible risk of mistakenly identifying them as extraterrestrial artifacts or probes. 

Previous work has used the Earth's shadow to search for real astrophysical transients lasting only a few seconds \citep{Richmond2020, Arimatsu2021, Nir2021}, but no significant findings have been reported thus far. Our objective, however, is to identify extraterrestrial artifacts moving in Earth's orbits and beyond.

Conducting searches in the Earth's shadow with a ground-based telescope permits the identification of {\it candidates} for active (self-luminous) extraterrestrial technological objects that are moving outside of the Earth's atmosphere, while a definitive identification will require follow-up triangulation and spectroscopic analysis for repeating or new events. We note that this approach may not be effective for detecting old or defunct ET artifacts in orbit around the Earth. This means that objects similar to those posited in \cite{Villarroel2022c} might not be detected; only objects with an intrinsic emission could be detected in this way. We note that the only intrinsic emission from known human objects are \textbf{mainly} from lasers designed for communications and from spacecraft propulsion systems (e.g., rocket thrusters); no other type of intrinsic emission from satellites is \textbf{widely} known.

\section{Searches for ZTF transients in Earth's Shadow}

The Zwicky Transient Facility \citep[ZTF;][]{Bellm2019,Graham2019} is an untargeted wide-field transient survey  observing transient events ranging from Solar System objects to powerful extragalactic events with the help of its systematic data acquisition system. The ZTF uses the 1.2m Samuel Oschin telescope at Palomar Observatory in California and reaches a limiting magnitude of $r \sim 20.5$.

The analyses in this section concern three sample sets of ZTF images.  These are labeled as samples A, B, and C.  Sample set A is defined and discussed in Section \ref{sec:setA}.  The much larger sample sets B and C are described in Section \ref{sec:independent}.  See Table \ref{tab_ztf_samples} for the definitions of each sample.

\begin{table*}
\caption{Definitions of ZTF image samples.}\label{tab_ztf_samples}
\begin{tabular}{|c|p{6cm}|c|p{6cm}|}
\hline
Sample & Definition & $N$ (images) & Purpose \\
\hline
A & Based on transient survey (courtesy of Igor Andreoni) and culled to include only images residing within 8$^\circ$ of the center of Earth's shadow $(\rho < 8^\circ)$& 678 & Demonstrating manual inspection method\\
B &  $(b > 20^{\circ}) \land (\rho < 6.5^{\circ})$ & 224,168 & Demonstrating automated method in largely uncrowded fields (avoiding low galactic latitudes ($\beta$)); located confidently within shadow ($\rho < 6$)\\
C & $\beta > 84^{\circ}$ & 326,823  & Control sample (ecliptic pole; ecliptic latitude ($\beta$) exceeds $84^{\circ}$; all fields well outside the shadow)\\
\hline

\end{tabular}
\end{table*}

\subsection{Examination of transient alerts (Sample A)}\label{sec:setA}

For sample A, we selected transient alerts from a public ZTF transient survey (courtesy of Igor Andreoni). The data set we searched spans from July 1, 2019, when real / bogus deep learning classification became available \citep{Duev2019}, until August 1, 2022. The data set excluded from the search those ZTF fields close to the Galactic plane to reduce contamination from M-dwarf stellar flares and other types of fast transients from astrophysical sources within the Milky Way. To achieve this, we ignored those fields with a Galactic extinction higher than E(B - V) = 0.3 mag at the central coordinates of the fields (using the \cite{Planck2014} dust maps) following \cite{Andreoni2020}.

One-off transients are, in principle, only visible in a single image before vanishing. This limits us to transients that, when bright and detectable (brighter than $\sim$ 20.5 mag), persist for a shorter time than the sum of the exposure time and the time difference with subsequent or previous images (this adds up to $\sim$ 30 seconds, since the subsequent image is captured in immediate succession).

Each alert included in the sample has at least 5 other alerts in the same image (i.e., same field, same JD) which were also detected only once during the ZTF survey, to focus on multiple transients. All sources within 1.5 arcseconds from a catalogued object were discarded. Also, all sources within 10 arcseconds from catalogued asteroids were removed. We obtained 11,029 transient alerts to examine closer.

\subsubsection{Exploiting the Earth's Shadow}

We use the Julian Date and the J2000 coordinates of each transient, as well as the software library {\tt earthshadow} published by Guy Nir published \citep{Nir2024_git}, in order to select only events that were observed in the shadow of the Earth at the altitude of geosynchronous orbits ($\rho < 8.5^{\circ}$ for Sample A; we use the stricter bounds of the DFS when filtering sets B and C: i.e., $\rho < 6^{\circ}$).

\begin{figure*}
   \includegraphics[scale=0.35]{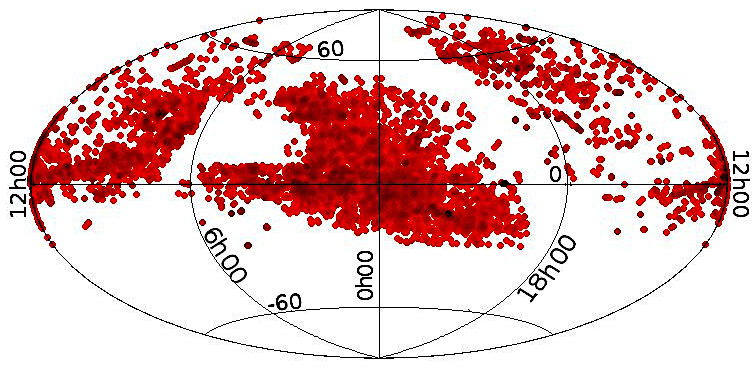}
   \\
   \includegraphics[scale=0.427]{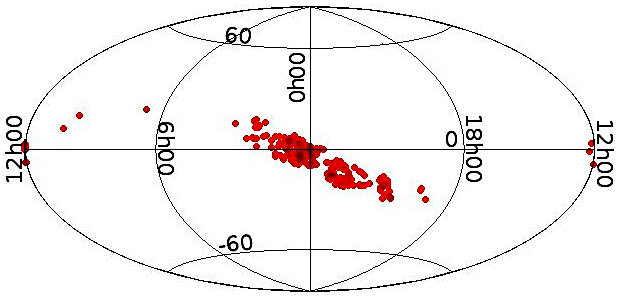}
  \caption{\label{psf2} {\bf Distribution on the sky.} The distribution of the Sample A transient candidates on the sky in Galactic coordinates. Upper panel (a) all one-off transient candidates, (b) one-off transient candidates in the shadow.}
   \end{figure*}

Out of the sample of 11029 candidates, 262 (2.4\%) of the examples are in the Earth's shadow, which excludes any specular reflections for these particular cases. We show the distributions of the candidate ``shadow transients'' over the sky in Figure \ref{psf2}.

\subsubsection{Classifying detections}

This exercise permitted us to quickly identify a set of classes of note: (1) images affected by a difference in depth (different magnitude limit) or other observational issues; (2) two images showing how a short transient is first visible in one image, and then another transient is seen in a second, later image (in nearly all cases of this kind, the transient is first seen in the eastern part of the image, and later in the western, which suggests a moving object); (3) images showing the synchronized movement of multiple objects moving from the east to west; (4) events that have been, in a preliminary sense, identified as candidate multiple transients. Category 1 objects are irrelevant to us. Category 2 objects are assigned a low priority as they are unlikely to be multiple transients, as indicated by their east-to-west movement. For Category 3, We find 29 cases of multiple objects moving throughout the field (from east to west), that we assume to be asteroids. Finally, in Category 4, we find 39 candidates of preliminary interest. However, examining them against the JPL Small-body Identification tool and a generous crossmatch radius of 30 arcseconds, identifies the majority of these as catalogued asteroids.

\begin{figure*}
   \includegraphics[scale=0.2]{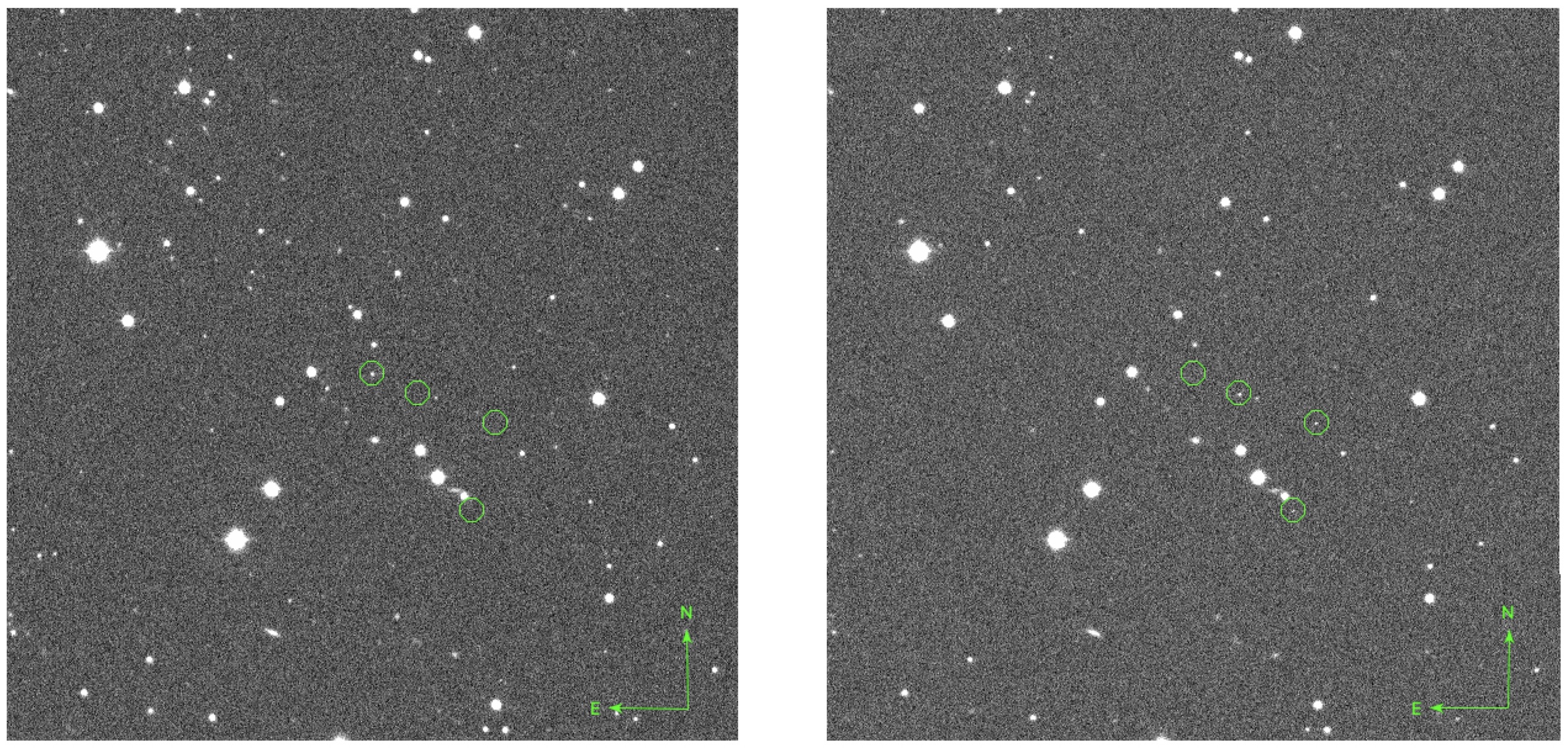}
  \caption{\label{tripletransient} {\bf Asteroid 2001 VC136.} A complex candidate, including asteroid 2001 VC136 and a transient was found inside a box measuring 10$\times$10 arcmin and centered at R.A. (J2000)=6.3634206 deg. and Dec.(J2000)= 
2.4341476, J.D.=2458760.7977431. See Table \ref{tab_tripletransient} for astrometric measurements and Table \ref{filenames} for data product filenames.  The lower-most object (right), not visible on the left, is a spurious detection (non-psf). }
   \end{figure*}

Two interesting candidates remain from Category 4. The first is the case of a possible transient appearing next to a moving asteroid, creating the impression of a ``triple transient,'' shown in Figure \ref{tripletransient}. The detections in the two left-most circles correspond to 2001 VC136, a $\sim$19.6 mag asteroid. The detections in the two right-most circles appear only once, although the lower turns out to be a spurious detection (two bright pixels, non-psf).

\begin{figure*}
   \includegraphics[scale=0.20]{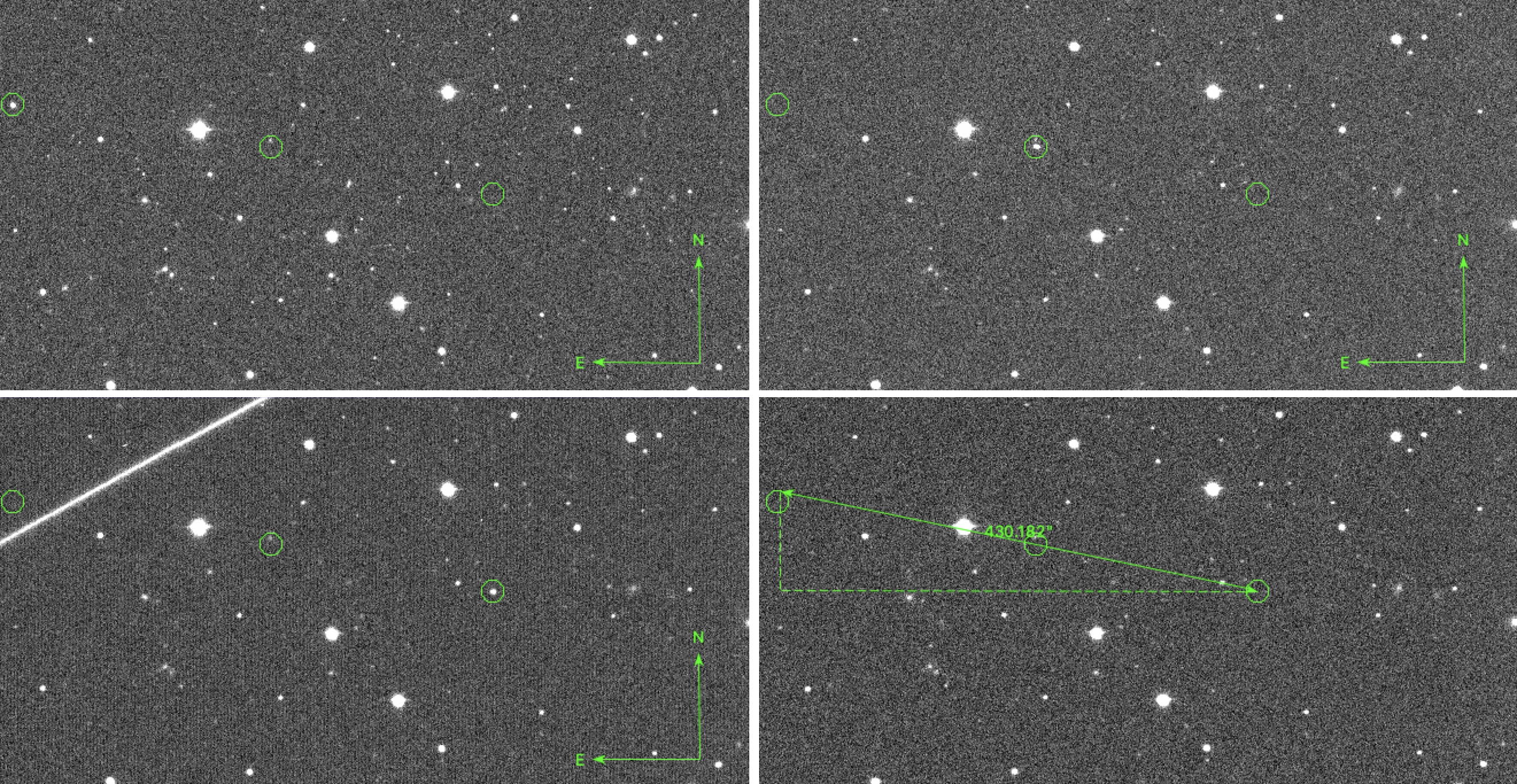}
  \caption{\label{weirdobject} {\bf Near-Earth object or triple transient?} The object(s) of interest can be found within a 1000 arcsecond box centered at R.A. (J2000)= 145.2448564 and Dec.(J2000)= 15.6224391, J.D.= 2459256.8096296; see Table 2 for astrometric coordinates. A luminous object is seen streaking through the upper-left corner of the third image (lower left), possibly a meteor or aircraft. A fourth image in the $g$ band with no object detected is added to the sequence (lower right), to show the deviation from a straight path. See Table \ref{UncataloguedAsteroid} for astrometric measurements and Table \ref{filenames} for data product filenames.}
   \end{figure*}

The second candidate is a good example of an uncatalogued object that could be interesting if more information were available.  Figure \ref{weirdobject} shows a $\sim$17 magnitude object or objects that have no match when queried using the JPL Horizons database in April 2025.  The astrometric measurements are tabulated in Table \ref{UncataloguedAsteroid}.  This object (or set of objects) occurs within $\sim$0.07 arcseconds $s^{-1}$ of the ecliptic, suggesting that it may be an asteroid.  On the other hand, if this sequence represents multiple images of the same object, then its apparent motion in right ascension is about 6$\times$ times greater than of typical main belt asteroids, suggesting it is closer to the Earth. The entire track spans 430 arcseconds during 100 minutes. The second set of images and positions ((b) and (c) in Table \ref{UncataloguedAsteroid}) were both acquired in the {\it g} band, and differ in maximum raw intensity above background by only 5\%.  Using these two positions, we predicted the expected location 71 minutes after the position in (c), assuming a linear path and constant angular speed (R.A. (J2000) = 145.163329$^{\circ}$, Dec. (J2000) = 15.608823$^{\circ}$) in the same field (570), filter ($g$), sensor (c02), and and quadrant (q3) with timestamp {\tt 20210211359132}: no object was visible because it was projected to reside in the seam gap between CCDs.  We also searched the neighboring sensor+quadrant in $r$ band images 136 minutes after (c), although the object was still projected to reside in the gap between quadrants and was not visible.  The image for (b) also appears somewhat streaked in the approximate direction of motion (East-West), although other objects in the scene (including stars) also appear to exhibit an elongation (if less pronounced) in the same orientation, suggesting an effect of inaccuracies in tracking.  

\begin{table*}
\caption{Astrometric measurements of the three transient candidates shown in Figure \ref{tripletransient}. Objects 1 and 2 were determined to be asteroid 2001 VC136. 
 Coordinates (R.A., Dec.) are given in J2000 for each detection, along with the Julian Date (J.D.) and UTC observation time. All observations were taken in the ZTF \textit{r}-band. Missing coordinate values are indicated as N.A.  ZTF data product filenames are listed in Table \ref{filenames}.\label{tab_tripletransient}}
\centering
\begin{tabular}{l c c c c c}
\hline\hline
\multicolumn{6}{c}{Asteroid 2001 VC136; see Figure \ref{tripletransient}} \\
\hline\hline
{\bf Object 1} & Band & R.A. (J2000) & Dec. (J2000) & J.D. & Observation time\\[3mm]
(a) left & \textit{r} & 6.3636953 & 2.4339307 & 2458760.7977431 & 2019-10-04 at 07:56:45.26\\
(b) right & \textit{r} & 6.3529988 & 2.4294331 & 2458760.8510417 & 2019-10-04 at 08:25:30.89 \\[3mm]
\hline\
{\bf Object 2} & Band & R.A. (J2000) & Dec. (J2000) & J.D. & Observation time\\[3mm]
(a) left & \textit{r} & N.A. & N.A. & 2458760.7977431 & 2019-10-04 at 07:56:45.26\\
(b) right & \textit{r} & 6.3355465 & 2.4229677 & 2458760.8510417 & 2019-10-04 at 08:25:30.89 \\[3mm]
\hline\
{\bf Object 3} & Band & R.A. (J2000) & Dec. (J2000) & J.D. & Observation time\\[3mm]
(a) left & \textit{r} & N.A. & N.A. & 2458760.7977431 & 2019-10-04 at 07:56:45.26\\
(b) right& \textit{r} & 6.3406132 & 2.4030027- & 2458760.8510417 & 2019-10-04 at 08:25:30.90 \\[3mm]
\hline\hline
\end{tabular}
\label{TripleMeasurements}
%\end{table*}
\end{table*}

\begin{table*}
\caption{Astrometric measurements of the uncatalogued object(s) shown in Figure \ref{weirdobject}. Coordinates (R.A., Dec.) are given in J2000 for each detection, along with the Julian Date (J.D.) and UTC observation time. Observations were taken on 2021 February 11 in the ZTF \textit{r} and \textit{g}-bands.  ZTF data product filenames are listed in Table \ref{filenames}.}
\centering
\begin{tabular}{l c c c c c}
\hline\hline
\multicolumn{6}{c}{Uncatalogued object(s) (Figure \ref{weirdobject})} \\
\hline\hline
\hline
{\bf Object} & Band & R.A. (J2000) & Dec. (J2000) & J.D. & Observation time\\[3mm]
(a) upper left & \textit{r} & 145.3668218 & 15.6441562 & 2459256.7390625 & 2021-02-11 at 05:44:16.26\\
(b) upper right & \textit{g} & 145.3011827 & 15.6335293 & 2459256.7789352 &  2021-02-11 at 06:41:40.693 \\
(c) lower left & \textit{g} & 145.2448868 & 15.6220271 & 2459256.8096296 &  2021-02-11 at 07:25:52.616 \\[3mm]
\hline\hline
\end{tabular}
\label{UncataloguedAsteroid}
%\end{table*}
\end{table*}

\subsubsection{Analysis of candidate clustered transients}

We performed an additional analysis by examining candidates that, in addition to appearing in images with $>5$ transients in the same field, two of these are within 10 arcminutes of each other. We found 85 cases of such clustered candidate transients. We then retrieved the images from the same day from the IRSA Gator's image survey for ZTF.\footnote{https://irsa.ipac.caltech.edu/applications/Gator/} We found that 23 of these cases have no transients at all visible in the image and 44 cases appear to be just an asteroid moving between two images. (Note this is hard to distinguish from the case of two transients happening close to each other (a double transient), which would look similar.) Two images have single transient candidates, and 7 show several probable asteroids moving in the same field.

Finally, we also conducted an investigation of the closest 10 arcminute field near each of the 262 transient candidates in the shadow. In this way, we found interesting cases of multiple asteroids moving together and one possible solitary transient.

We note that, with only 2 or 3 exposures of 30 seconds on each field every day, it is difficult or unlikely to find multiple transients if they happen over half an hour.
We also note that the hypothetical case of a double transient might be indistinguishable from an asteroid if they appear in only two images, and might only be distinguished by the fact that asteroids move always from east to west.

Groups or clusters of objects moving in similar ways through a scene are unusual and potentially of interest for a study like the one proposed here.  The greatest likelihood is that these represent (1) clusters of asteroids or (2) clusters of terrestrial spacecraft or spacecraft components in heliocentric orbit.  Alternatively, such detections could represent (3) clusters of artificial objects of extraterrestrial origin in heliocentric orbit or moving in formation.  Reflectance spectra are likely essential for distinguishing these cases if the inferred kinematics are prosaic.

\begin{figure*}
   \includegraphics[scale=1.5]{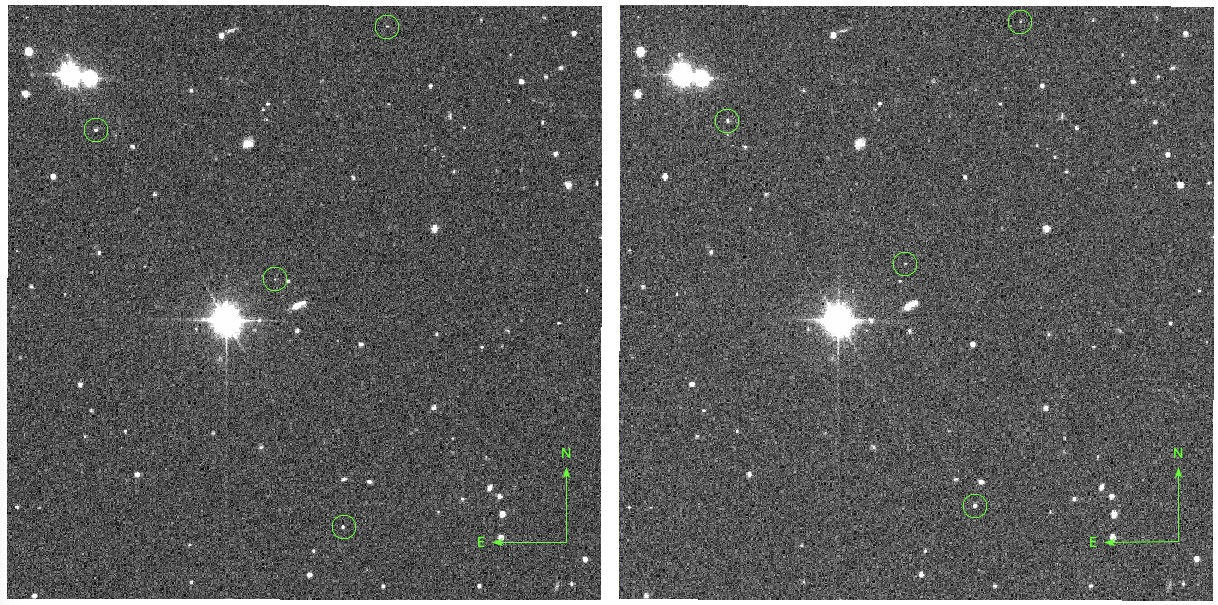}
  \caption{\label{example1} {\bf Example with multiple asteroids.} From uppermost to lowermost in each image, these objects are the catalogued asteroids 237629 (2001 RZ128), 34921 (4801 P-L), 468722 (2010 GF35), and 408570 (2013 LO5), respectively; they are shown here moving between these image, captured 34 min apart. Both images are $r$ band images and were observed on 2021-03-19.  See Table \ref{Measurements1} for astrometric parameters, and Table \ref{filenames} for ZTF data product filenames.}
   \end{figure*}

\begin{figure*}
   \includegraphics[scale=0.4]{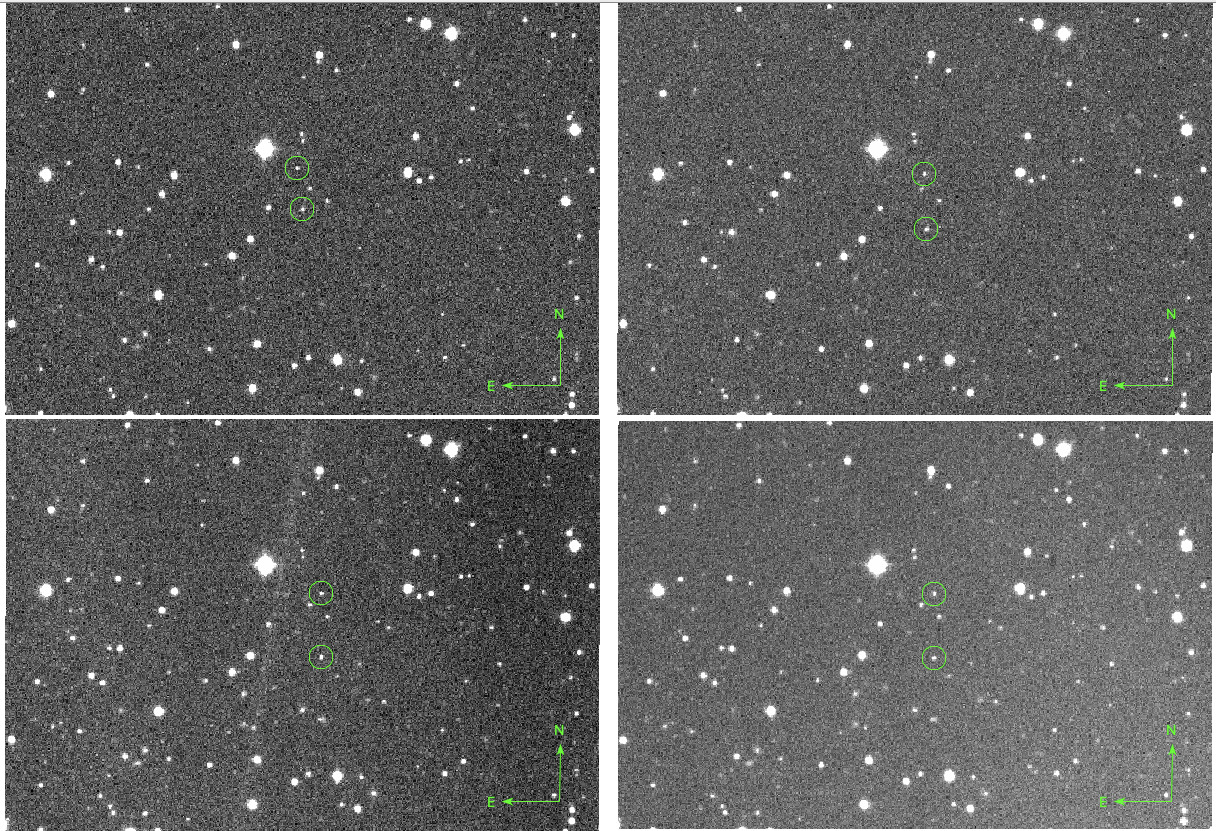}
  \caption{\label{example2} {\bf Example with multiple asteroids.} These objects were captured within $3^{\circ}$ of the Earth shadow center. The two upper images are $g$ band and the two lower images are $r$ band. The uppermost and lowermost objects within each image have been identified as the asteroids 239154 (2006 KX9) and 728206 (2010 LF127), respectively.  See Table \ref{Measurements2} for astrometric measurements. ZTF data product filenames are listed in Table \ref{filenames}.}
   \end{figure*}

The first two clusters were observed  crossing a region 10 by 10 arcminutes in size, at angular velocities of $\sim$1 arcsecond per minute. We show two examples of fields with multiple asteroids moving from east to west (see Figures \ref{example1} and \ref{example2}). Figure \ref{example1} shows four objects moving in two frames, each with a normal 30s ZTF exposure, while Figure \ref{example2} shows two objects moving across four frames. The individual coordinates for the two examples can be found in Tables \ref{Measurements1} and \ref{Measurements2}. The five other cases (of the 7 earlier mentioned) were observed on 2019-07-29, 2019-09-23, 2019-10-04, 2020-10-08, and 2021-09-02.  All told, each of the seven cases were found to involve multiple asteroids moving in tandem (similar angular speeds) from east to west.  In some cases, the displacements are of comparable size during the same time interval, but the direction of the motion is not exactly the same for each object in a given image, suggesting that the observation records a crossing of orbits of comparable radius.

We investigated whether any of these objects are found in the JPL Horizon databases. Indeed, both cases illustrated in Figures \ref{example1} and \ref{example2} are catalogued asteroids. Every single case of multiple asteroids that we have found in Sample A are of previously identified asteroids.  

\begin{table*}
\caption{Astrometric measurements of three moving objects identified in Figure \ref{example1}. Coordinates (R.A., Dec.) are given in J2000 for each detection, along with the Julian Date (J.D.) and UTC observation time. All observations were made on 2021 March 19 in the SDSS \textit{r}-band.  ZTF data product filenames are listed in Table \ref{filenames}.  Objects 1 to 4 have been identified as asteroids 237629 (2001 RZ128), 34921 (4801 P-L), 468722 (2010 GF35), and 408570 (2013 LO5), respectively.}
\centering
\begin{tabular}{l c c c c c}
\hline\hline
\multicolumn{6}{c}{Example case 1} \\
\hline
\hline
{\bf Object 1} & Band & R.A. (J2000) & Dec. (J2000) & J.D. & Observation time \\[3mm]
(a) left & \textit{r} & 182.2828716 & 2.3212253 & 2459292.8540046 & 2021-03-19 at 08:29:46.685 \\
(b) right & \textit{r} & 182.2779462 & 2.3223948 & 2459292.8789236 & 2021-03-19 at 09:05:40.534 \\[3mm]
\hline
{\bf Object 2}  &  &  &  &  &  \\[3mm]
(a) left & \textit{r} & 182.3511210 & 2.2971117 & 2459292.8540046 & 2021-03-19 at 08:29:46.685 \\
(b) right & \textit{r} & 182.3466646 & 2.2992191 & 2459292.8789236 & 2021-03-19 at 09:05:40.534 \\[3mm]
\hline
{\bf Object 3} &  &  &  &  &  \\[3mm]
(a) left & \textit{r} & 182.3091393 & 2.2621854 & 2459292.8540046 & 2021-03-19 at 08:29:46.685 \\
(b) right & \textit{r} &  182.3049176 & 2.2656986 & 2459292.8789236 & 2021-03-19 at 09:05:40.534 \\[3mm]
\hline
{\bf Object 4} &  &  &  &  &  \\[3mm]
(a) left & \textit{r} & 182.2933499 & 2.204017 & 2459292.8540046 & 2021-03-19 at 08:29:46.685 \\
(b) right & \textit{r} & 182.2885649 & 2.209137 & 2459292.8789236 &  2021-03-19 at 09:05:40.534\\
%\multicolumn{6}{c}{Date of observation= 2021-03-19} \\[3mm]
\hline\hline
\end{tabular}
%\end{table*}
\label{Measurements1}
\end{table*}

\begin{table*}
\caption{Astrometric measurements of two moving objects shown in Figure \ref{example2}. Coordinates (R.A., Dec.) are given in J2000 for each detection, along with the Julian Date (J.D.) and UTC observation time. All observations were made on 2019 July 29 in the SDSS \textit{g}- and \textit{r}-bands.  ZTF data product filenames are listed in Table \ref{filenames}.  Objects 1 and 2 have been identified as the asteroids 239154 (2006 KX9) and 728206 (2010 LF127), respectively.}
\centering
\begin{tabular}{l c c c c c}
\hline\hline
\multicolumn{6}{c}{Example case 2} \\
\hline
\hline
{\bf Object 1} & Band & R.A. (J2000) & Dec. (J2000) & J.D. & Observation time\\[3mm]
(a) upper left & \textit{g} & 309.9298057 & -26.9091492 & 2458693.8131481 & 2019-07-29 at 07:30:56.985 \\
(b) upper right & \textit{g} & 309.9258638 & -26.9105563 & 2458693.8291319 & 2019-07-29 at 07:53:57.595 \\
(c) lower left & \textit{r} & 309.9234980 & -26.9112611 & 2458693.8384954 & 2019-07-29 at 08:07:27.393 \\
(d) lower right & \textit{r} & 309.9232354 & -26.9114958 & 2458693.8393981 & 2019-07-29 at 08:08:45.309 \\[3mm]
\hline
{\bf Object 2}  &  &  &  &  &  \\[3mm]
(a) upper left & \textit{g} & 309.9284917 & -26.9187663 & 2458693.8131481 & 2019-07-29 at 07:30:56.985 \\
(b) upper right & \textit{g} & 309.9253378 & -26.9234575 & 2458693.8291319 & 2019-07-29 at 07:53:57.595 \\
(c) lower left & \textit{r} & 309.9234978 & -26.9262724 & 2458693.8384954 & 2019-07-29 at 08:07:27.393 \\
(d) lower right & \textit{r} & 309.9232349 & -26.9265072 & 2458693.8393981 & 2019-07-29 at 08:08:45.309\\[3mm]
%\multicolumn{6}{c}{Date of observation= 2019-07-29} \\[3mm]
\hline
\hline
\end{tabular}
%\end{table*}
\label{Measurements2}
\end{table*}

The searches described in this section for ZTF transients with durations shorter than 30 seconds in the Earth's shadow resulted in only one detection of significant interest, which may represent an uncatalogued asteroid (see Table \ref{UncataloguedAsteroid} and Figure \ref{weirdobject}).  This work illustrates a manual approach to evaluating in-shadow candidate transients (in this case identified by a 3rd party automated search) as candidate exotic objects, including artifacts of ET origin.  In the next section, we demonstrate a relatively exhaustive and automated search using image processing techniques to search for non-catalogue (``unmatched'') objects in the Earth's shadow.

\subsection{Automated transient survey (sets B--C)}\label{sec:independent}

Sample sets B and C, which are derived from the set of all publicly-accessible ZTF images predating 2024-04-15, are defined in Table \ref{tab_ztf_samples}.  Sample B was confined to the Earth's shadow and designed to avoid glinting satellites.  In particular, sample B is confined to an area within 6.5$^{\circ}$ of the center of Earth's shadow (i.e., $\rho < 6.5^{\circ}$, for search radius $\rho$ from the shadow center), avoiding the most crowded fields at low galactic latitudes ($b>20^{\circ}$). Sample C is a control sample, residing completely outside of the Earth's shadow, encompassing a 6$^\circ$ radius from the north ecliptic pole, where few to no minor bodies are expected ($\beta > 84^{\circ}$).

The goal of our search was to identify nonastrophysical transients unrelated to satellite glints. Of special interest were luminous objects that are (i) markedly streaking; (ii) point sources that cannot be identified as planetesimals in heliocentric orbit, (iii) clustered transients, and especially (iv) trains of point sources, as of flashing, luminous objects or of metallic glinting objects from beyond geosynchronous altitudes.  The prospect of detecting objects that meet one of these descriptions can be used to motivate triangulation for estimating range, in order to determine whether an object of interest resides within Earth's atmosphere and, if not, in order to estimate its orbital parameters.  This could permit detailed follow-up observations and further characterization.

We detected transient candidates using the segmentation method in the {\tt photutils}  library\citep{Bradley2024}. For each ZTF FITS image in the sample, the following procedure was carried out using a software package called {\it NEOrion} that was developed in-house.\footnote{Neorion is an ancient Greek word for ``shipyard''; Orion was a hunter; NEOs are Near Earth Objects.} 

\begin{enumerate}

\item {\bf Detection:} A detection threshold of 3$\sigma$ was applied above an adaptive (spatially varying) estimate of the background, essential for nonflat fields.  The minimum area for segment detection was 9 connected pixels.

\item {\bf Characterization:} We calculated and stored parameters related to
  intensity and morphology, to identify artifacts (e.g., cosmic ray tracks) and to identify objects exhibiting specific features (e.g., elongation/streaking).  

  \begin{itemize}

  \item S/N ($q$): $\mathcal{F}/(\sigma A)$, where $\mathcal{F}$ is the integrated flux and $\sigma$ is the standard deviation of image intensities for a 3-sigma clipped copy of the image. 
  
  \item Orientation ($\psi$): Angle of semimajor axis orientation ($\pm 90^{\circ}$), where $0^{\circ}$ corresponds to horizontal.

  \item Length (semimajor deviation, $\sigma_a$): one-sigma standard deviation along the semi-major axis of a 2D Gaussian w/same 2nd-order moments as the source.
    
  \item Width (semiminor deviation, $\sigma_b$): one-sigma standard deviation along the semi-minor axis of a 2D Gaussian w/same 2nd-order moments as the source.
    
  \item Elongation ($\eta$): $\sigma_a/\sigma_b$ \item Ellipticity ($\varepsilon$): $(\sigma_a-\sigma_b)/\sigma_a$; near zero for radially-symmetric objects
    
  \item Area ($A$): Area of the segment in pixels.  \item Equivalent radius ($R$): Radius of a circle of area $A$.  
  
  \item Gini coefficient ($G$): Gini coefficient of intensities in the segment.  This is near 0 for uniform illumination and near 1 if brightness is concentrated in a few pixels.  \item Maximum intensity ($p$): Maximum intensity of pixels inside the segment.  (The Gini coefficient has applications studies of galaxy morphology: see e.g., \cite{lisker2008gini}.)
  
  \item Taper ($\Upsilon$): We define the ``taper'' as the ratio of the FWHM of the minor axis of the segment ($\sim 2.35\times \sigma_b$) to the ``seeing FWHM'' reported in the FITS meta data.  For point sources and straight-line streaking point sources, $\Upsilon \sim 1$.
    
  \end{itemize}

For all detections, we also tabulated $d_{\rm edge}$, the distance in pixels from the edge of the field, where many linear artifacts tend to reside.
  
\item {\bf Labeling unmatched objects}: object centroids were estimated and then
  the angular distances calculated to the nearest object in the Pan-STARRS (Panoramic Survey Telescope and Rapid Response System, catalog PS1 DR2; \cite{kaiser2002pan,flewelling2020pan}) and SDSS (Sloan Digital Sky Survey; cite{york2000sloan}) catalogs residing in the field-of-view ($d_{\rm cat}$).  As discussed later, a threshold is applied to $d_{\rm cat}$ to identify ``unmatched'' objects.

\item {\bf Counting revisits}: We tallied the number of times ZTF captured an
  image (of each filter type) that overlapped the position of each unmatched detection ($n_{\rm r}, n_{\rm g}, n_{\rm b}$).  This was used later to filter out transient detections whose positions were captured only a few times.  For example, the celestial coordinates of an unmatched detection having $n_{\rm g} = 7$ was captured 7 times by the {\it g} filter. A detection with a low revisit count may simply be a variable astronomical source that happened to dim below the detection threshold.  A detection with a high revisit count and no proximate neighbors is more likely to be a genuine transient.

\item {\bf Computing nearest neighbor distances}: We computed the nearest
  neighbor distance between each unmatched detection and all of the others ($d_{\rm nn}$).  This was used later to filter out false positives: i.e., objects that could not be matched to catalog objects, but which appeared in multiple images at the same (or very proximate) location: these are likely to be slow-moving or unmapped/uncataloged sources.  An undesirable consequence is that tightly-clustered transients are eliminated in this way.

\item {\bf Filtering}: For ZTF samples B and C, we filtered the unmatched detections
  to create several subsets.  The goal of the first filter pass (producing set I) was to identify all valid detections.  The goal of the second filter pass (producing set II) was to remove likely artifacts like sensor noise and cosmic ray tracks.  The third filter pass (set III) preserved bright, isolated objects.  The fourth filter pass was used to identify elongated (streaking) objects.  The conditions used to define each subset are defined in Tables \ref{tab_conditions} and \ref{tab_subsets}.  Note that $c_7$ is not used in the definition of subsets for the control sample C.  In our discussion of results, we refer to the specific subset of detections obtained using a specific method from a specific ZTF sample as follows: {\tt sample/subset}.  For example, subset II of ZTF sample C is denoted ``C/II''.  Note that we have also disregarded all images with INFOBITS $\geq 33554432$, which ZTF has used to mark invalid data products.

\end{enumerate}

\begin{table*}
\caption{Conditions used to filter unmatched detections from ZTF samples B and C into subsets; see also Table \ref{tab_subsets}.}
\label{tab_conditions}
\begin{tabular}{|c|c|p{8.5cm}|}
\hline\hline
Condition & Definition & Purpose \\
\hline\hline
$c_1$ & $p < 3\times 10^4$ & not saturated\\
$c_2$ & $d_{\rm nn} > 10$ as & separation from nearest unmatched detection\\
%$c_2^{\prime}$ & $d_{\rm nn} > 5$ as & separation from nearest unmatched detection\\
%$c_2^{\prime\prime}$ & $d_{\rm nn} > 20$ as & separation from nearest unmatched detection\\
$c_3$ & $d_{\rm cat} > 5$ as & separation from nearest catalog objects \\
$c_3^{\prime}$ & $d_{\rm cat} > 20$ as & highly separated from nearest catalog objects \\
$c_4$ & $d_{\rm edge} > 20$ pix & at least 20 pixels from edge\\ 
$c_5$ & $(n_{\rm r} > 3) \lor (n_{\rm g} > 3) \lor (n_{\rm b} > 3)$ & location visited at least 10 times\\
$c_6$ & $q > 5$ & high S/N \\
$c_6^{\prime}$ & $q > 10$ & very high S/N \\
$c_7$ & $\rho < 6^{\circ}$ & within dark full shadow (DFS) \\
$c_8$ & $(\varepsilon < 0.15) \lor ((|\psi| > 5^{\circ}) \land (|\psi| < 85^{\circ}))$ & not a vertically- or horizontally-oriented elongated object (typical of many artifacts caused by telescope exposure and tracking errors)\\
$c_9$ & $G > 0.1$ & removes uniform-intensity artifacts\\
$c_9^{\prime}$ & $G > 0.2$ & removes uniform-intensity artifacts\\
$c_9^{\prime\prime}$ & $G > 0.5$ & removes uniform-intensity artifacts\\
$c_{10}$ & $10 < A < 1000$ & removes tiny and very large objects\\
$c_{11}$ & $\eta > 10 $ & elongated objects (aspect ratio larger than 10) \\
%$c_{11}^{\prime}$ & $\eta > 10 $ & symmetric objects (aspect ratio less than 2) \\
%$c_{20}$ & $s < 0.8$ & removes most cosmic ray tracks\\
%$c_{20}^{\prime}$ & $s < 0.5$ & unlikely for point source (resolved)\\
$c_{12}$ & $ 0.3 < \Upsilon < 1.3$ & removes cosmic rays and uniform-intensity artifacts\\
$c_{12}^{\prime}$ & $ 0.3 < \Upsilon < 1$ & removes cosmic rays and uniform-intensity artifacts\\
\hline\hline
\end{tabular}
\end{table*}

\begin{table*}
\caption{Definitions of subsets of unmatched detections.}\label{tab_subsets}
\begin{tabular}{|c|c|c|c|p{6cm}|}
\hline\hline
Subset & Definition & $N$ detections (Sample B) & Description\\
\hline\hline
I   & $c_3 \land c_5 \land c_7$ & 3,734,423 & All valid dections\\
% set 1 in analyze_unmatched_seg.py used for xnz4:
II  & $c_1 \land c_3  \land c_4 \land c_5 \land c_7 \land c_6        \land c_8  \land c_9  \land c_{10} \land c_{12}$ & 614,556 & Most artifacts removed\\
% set 2 in analyze_unmatched_seg.py used for xnz4:
III & $ c_1 \land c_2 \land c_3^{\prime} \land c_4 \land c_5 \land c_6^{\prime} \land c_7 \land c_8 \land c_9^{\prime} \land c_{10} \land  \land c_{12}^{\prime}$ & 16,693 & Bright and isolated \\
% set 4 in analyze_unmatched_seg.py used for xnz4:
IV  & $c_1 \land c_2 \land c_4 \land c_5 \land c_6 \land c_7 \land c_8 \land c_{11} \land c_{12}^{\prime}$ & 5,940 & Bright and elongated \\
\hline\hline
\end{tabular}
\end{table*}

\begin{figure*}
   \includegraphics[scale=0.38]{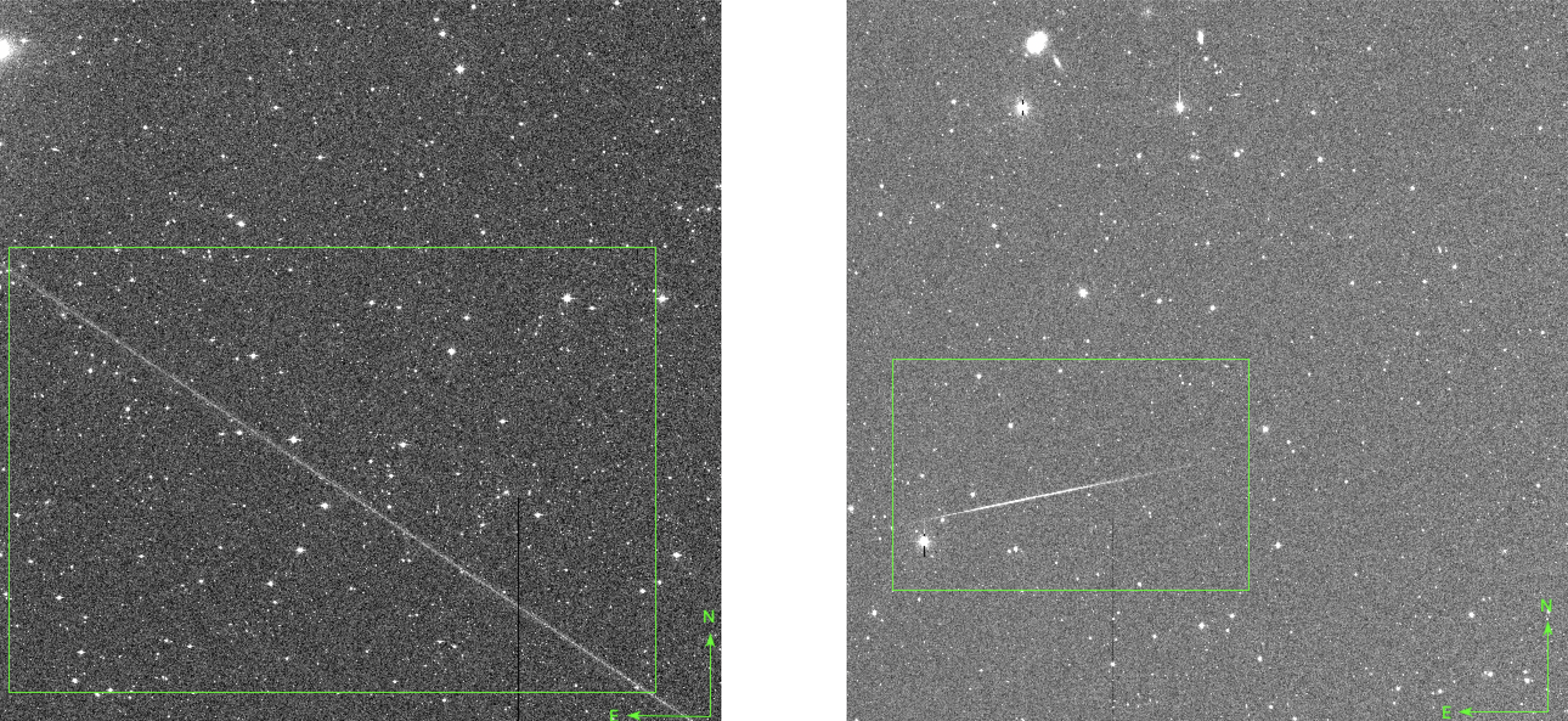}
   \\
   \includegraphics[scale=0.4]{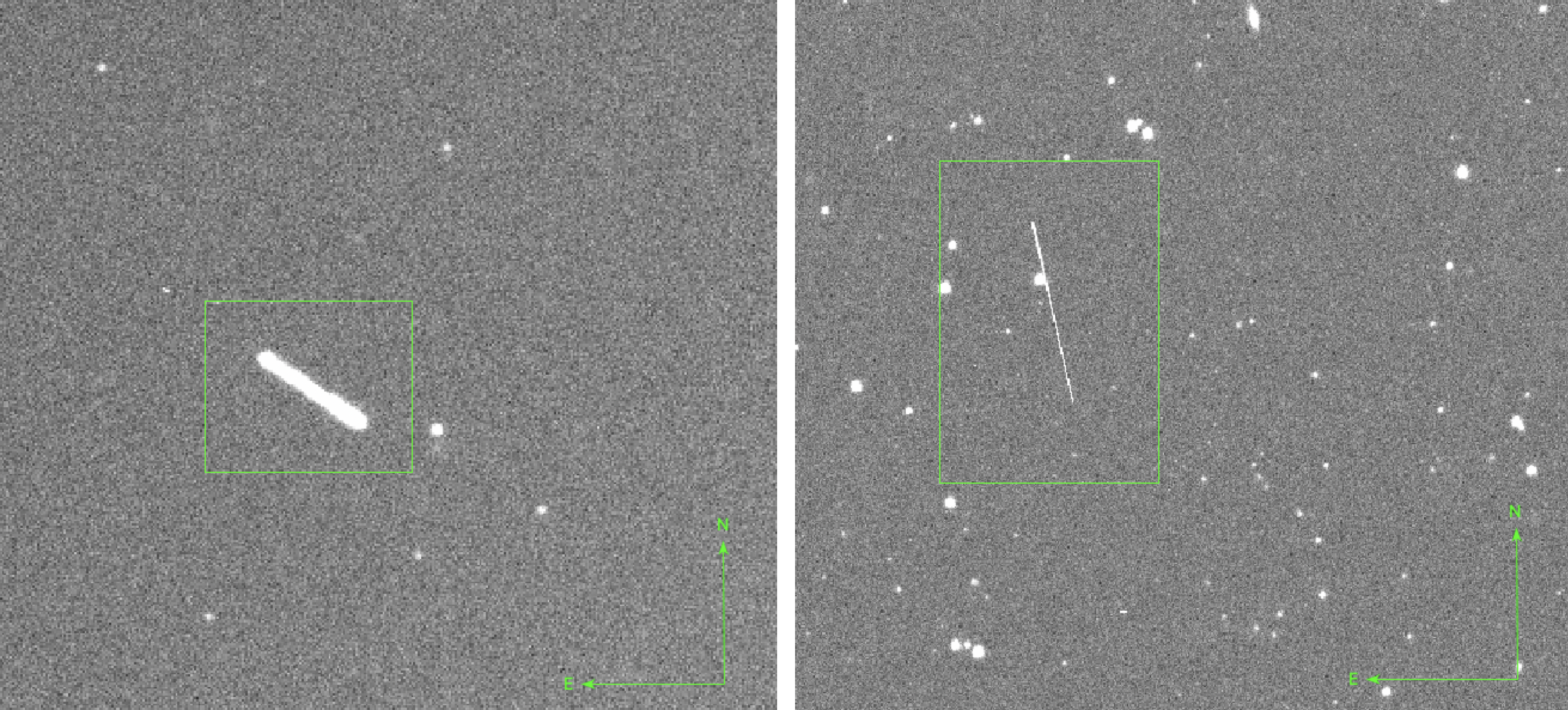}
  \caption{{\bf Examples of streaks found near the shadow center, from B/IV.} The box sizes in the upper panels are (a) 46.3 $\times$ 31.9 arcmin, and (b) 25.2 arcmin $\times$ 16.6; in the lower panels: (c) 1.7 $\times$ 1.4 arcmin and (d) 2.6 $\times$ 3.8 arcmin. The first three images (a) through (c) are believed to be observations of real objects, while the fourth (d) is likely a cosmic ray due to the narrowness (FHWM of $\sim$1 arcsecond) and uniform width of the track in spite of brightness fluctuations. Given the 30 second exposure time and assuming that the objects are gravitationally bound to Earth, the object in (a) is moving at angular speeds consistent with orbits lower than $<$ 14,000 km, while object (b) corresponds to $\sim$ 20,000 km altitude.  Both objects are instead probably meteors (see text). (c) This uncatalogued object (not in MPC database as of April 2025) was 4.5$^{\circ}$ from the shadow center, and is probably an unknown heliocentric asteroid. If instead this were in a circular geocentric orbit, the streak length implies an altitude of almost 200,000 km, where the shadow radius is only about 2$^{\circ}$.  Refer to Table \ref{filenames} for the ZTF data product filenames.}\label{coolstreaks}
   \end{figure*}

\begin{figure*}
   \includegraphics[scale=0.47]{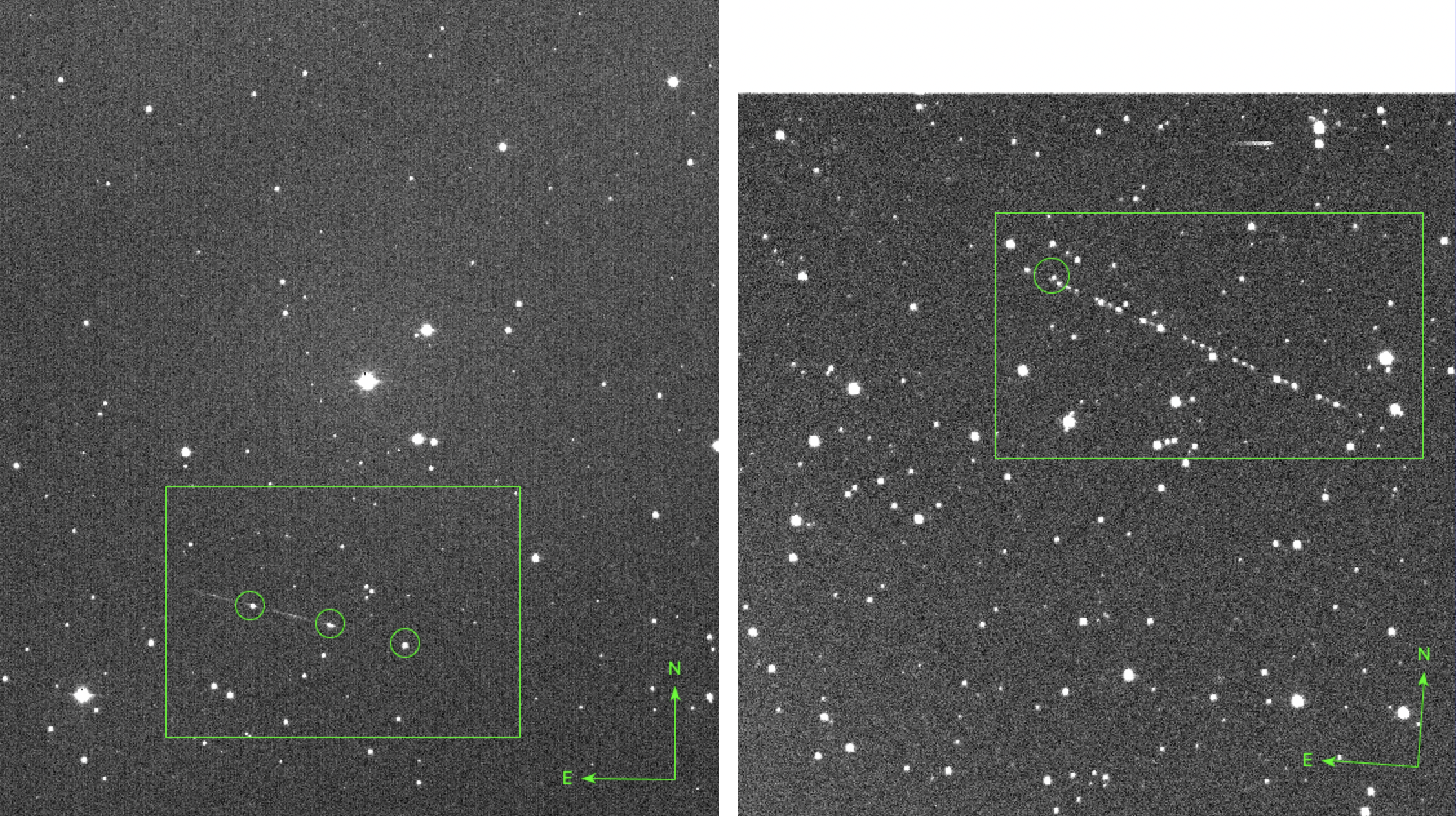}
   \includegraphics[scale=0.47]{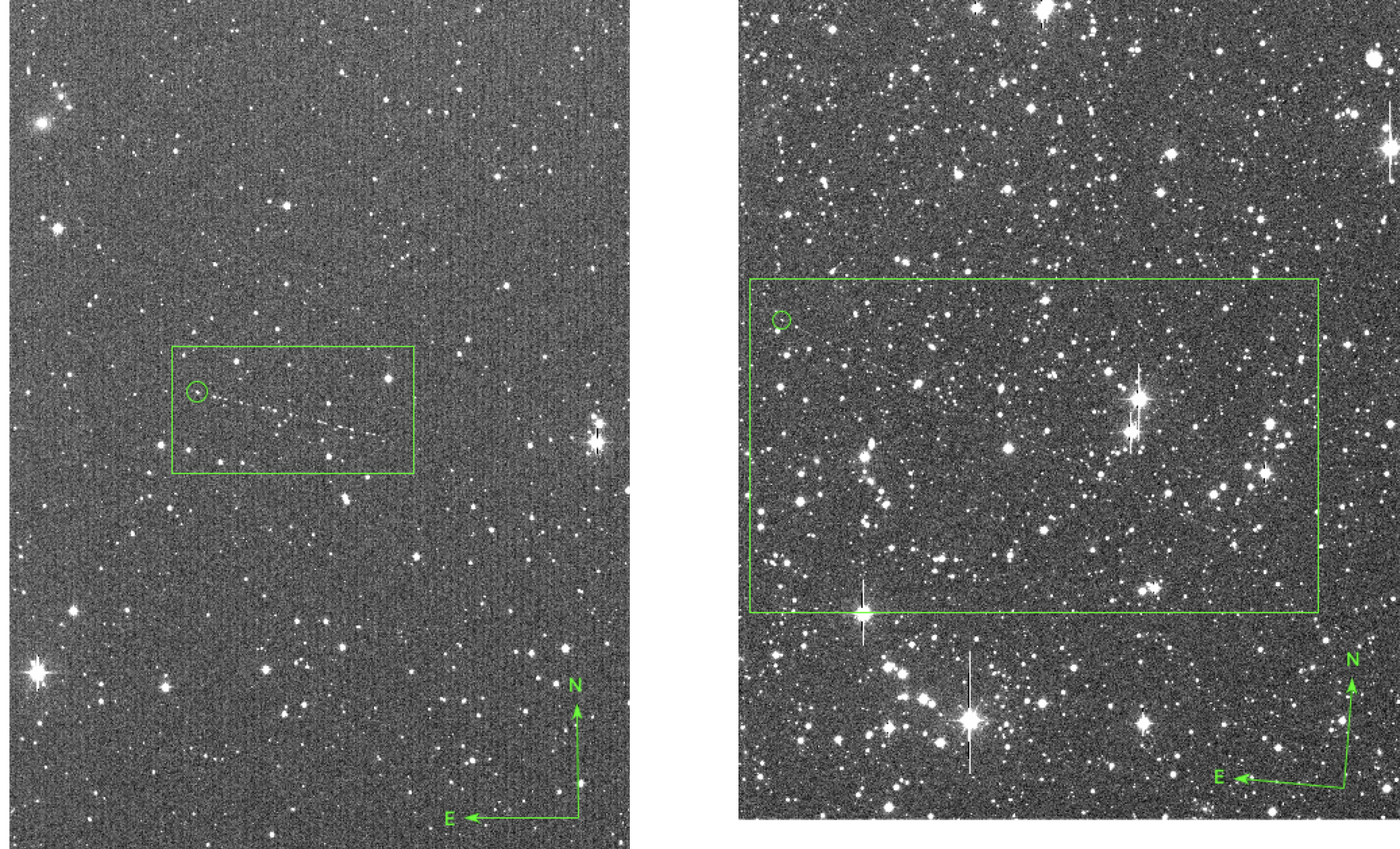}
   \caption{{\bf Examples of flash trains caused by satellite glints near the ecliptic pole.} The box sizes for the upper panels are (a) 8.6 $\times$ 5.7 arcmin, and (b) 8 arcmin $\times$ 4.6; for the lower panels: (c) 7.5 $\times$ 3.9 arcmin, and (d) 21.2 $\times$ 12.4 arcmin.  Given the 30 second exposure time and assuming that the objects are gravitationally bound to Earth, these objects are moving at angular speeds consistent with medium Earth orbits to geosynchronous orbits.  Refer to Table \ref{filenames} for the ZTF data product filenames.}\label{Ecliptic}
   \end{figure*}

For the present proof-of-concept study, we manually reviewed the detections to determine what kinds of objects do and do not appear in these samples, rather than produce an exhaustive catalogue of the results.  For example, in Figure \ref{coolstreaks} we present a few examples from B/IV of streaking objects. Streaks resembling these may belong to one of several categories: (i) meteors; (ii) high-altitude aircraft lights; (iii) near-Earth asteroids located beyond the shadow cone; or (iv) light-emitting objects above the atmosphere and moving within the shadow cone. The estimated angular speeds based on streak length, exposure time (30 seconds), and the assumption of continuous emission, can be used to estimate altitude in the case of objects in circular orbits around Earth.  This can be used estimate whether, granting these assumptions, they reside in the shadow cone.   Confident assignment to (iv) is impossible, however, without estimating a parallax, which is a primary objective of planned future searches using ExoProbe \citep{ExoProbe2023}.  

``Flash trains'' are linear alignments of point source flashes, sometimes with intermediate streaking, commonly associated with satellites but also conceivably produced by pulsing light sources.  NEOrion found no flash trains in Sample B (in shadow).  By contrast, flash trains were readily found in Sample C (near the ecliptic pole); we show typical examples in Figure \ref{Ecliptic}.  These exhibit angular speeds consistent with altitudes typical of Medium Earth Orbit (MEO) to geosynchronous orbit (GSO).

NEOrion was also able to find clusters of previously catalogued minor bodies, demonstrating the ability to find clusters of transient candidates in single images.  Manual follow-up inspection was required to find displaced images of the same objects in time-adjacent images, in order to recognize them as planetesimals.  For example, we found four previously catalogued minor bodies in a single ZTF image; these are listed in Table \ref{4asteroids}. 
To find multiple transient candidates similar to those presented in \cite{Villarroel2021,Solano2024}, however, one needs to search for clusters of transients in time-adjacent images corresponding to approximately one hour of total exposure time. We have so far not automated this follow-up confirmation process but have plans to do so in future: that is, to download temporally adjacent images of the same field and then search for comparably bright objects that have been displaced in a westward direction.  This enhancement would also permit us to search for candidate transients that are clustered in time.

\begin{table}
\caption{Cluster of previously catalogued minor bodies detected using the automated search in the shadow, Sample B (all objects were found within 5.7$^{\circ}$ of the Earth shadow center).  All were found in {\tt ztf\_20230917395197\_000445\_zr\_c04\_o\_q4\_sciimg.fits}, captured at 2023-09-17T09:29:06.489 (J.D. 2460204.8951968).  Displaced images of these objects were found in temporally adjacent scenes.} 
\label{4asteroids}
\centering
\begin{tabular}{c c c}
\hline\hline
\multicolumn{3}{c}{Cluster of four catalogued minor bodies} \\
\hline\hline
R.A. (J2000) & Dec. (J2000) & Object ID\\
\hline
345.58864 & -5.91241 & 6459 Hidesan (1992 UY5) \\
345.19842 & -5.78249 & 18555 Courant (1997 CN4) \\
345.50169 & -5.55202 & 29453 (1997 RU6)\\
345.68042 & -5.62310 & 625124 1997 GK23)\\
\hline\hline
\end{tabular}
%\end{table*}
\end{table}

\begin{table*}
\caption{Table of filenames used in figure components, with distance from center of detected objects to center of Earth shadow.}
\label{filenames}
\begin{tabular}{|c|l|p{1.25cm}|c|}
\hline\hline
Figure & Component & Shadow center distance & FITS filename \\
\hline\hline
\ref{tripletransient} & (a) left & $4.6^{\circ}$ & ztf\_20191004297708\_000447\_zr\_c02\_o\_q1\_sciimg \\
  & (b) right & $3.0^{\circ}$ & ztf\_20191004351042\_000447\_zg\_c02\_o\_q1\_sciimg \\
\hline
\ref{weirdobject} & (a) upper left & $5.9^{\circ}$ & ztf\_20210211239062\_000570\_zr\_c02\_o\_q3\_sciimg \\
  & (b) upper right & $4.8^{\circ}$ & ztf\_20210211278912\_000570\_zg\_c02\_o\_q3\_sciimg \\
  & (c) lower left & $4.3^{\circ}$ & ztf\_20210211309618\_000570\_zg\_c02\_o\_q3\_sciimg \\
  & (d) lower right & $4.5^{\circ}$ & ztf\_20210211359132\_000570\_zg\_c02\_o\_q3\_sciimg\\
\hline
\ref{example1} & (a) left & $7.5^{\circ}$ & ztf\_20210319353982\_000473\_zr\_c04\_o\_q2\_sciimg\\
  & (b) right & $8.2^{\circ}$ & ztf\_20210319378912\_000473\_zr\_c04\_o\_q2\_sciimg\\
\hline
\ref{example2} & (a) upper left & 1.0$^{\circ}$ & ztf\_20190729312951\_000287\_zg\_c02\_o\_q2\_sciimg \\
  & (b) upper right & 1.6$^{\circ}$ & ztf\_20190729329120\_000287\_zg\_c02\_o\_q2\_sciimg \\
  & (c) lower left & 2.1$^{\circ}$ & ztf\_20190729338495\_000287\_zr\_c02\_o\_q2\_sciimg \\
  & (d) lower right & 2.8$^{\circ}$ & ztf\_20190729339398\_000287\_zr\_c02\_o\_q2\_sciimg \\
\hline
\ref{coolstreaks} &  (a) upper left & 4.9$^{\circ}$ & ztf\_20190210285995\_000467\_zr\_c09\_o\_q1\_sciimg\\
  & (b) upper right & 4.4$^{\circ}$ & ztf\_20191027260613\_000504\_zr\_c16\_o\_q4\_sciimg\\
  & (c) lower left & 4.5$^{\circ}$ & ztf\_20191011385475\_001494\_zg\_c03\_o\_q4\_sciimg\\
  & (d) lower right & 3.3$^{\circ}$ & ztf\_20180912341435\_000394\_zi\_c02\_o\_q1\_sciimg\\
\hline
\ref{Ecliptic} & (a) upper left & 79.4$^{\circ}$ & ztf\_20190326382303\_000825\_zg\_c10\_o\_q2\_sciimg\\
  & (b) upper right & 85.2$^{\circ}$ & ztf\_20190326515729\_000825\_zr\_c09\_o\_q2\_sciimg\\
  & (c) lower left & 85.8$^{\circ}$ & ztf\_20190408450498\_000826\_zr\_c11\_o\_q1\_sciimg\\
  & (d) lower right & 81.8$^{\circ}$  & ztf\_20190411421551\_000825\_zr\_c09\_o\_q3\_sciimg\\
\hline\hline
\end{tabular}
\end{table*}
   
\section{Discussion}

In this paper, we described techniques that can be used to search for extraterrestrial artifacts in the Solar System near Earth. We discussed the utilization of pre-Sputnik images, space-borne telescopes, the analysis of reflectance spectra of space debris, and using the Earth's shadow as a filter to facilitate the detection of such objects. Previous studies \citep{Villarroel2022b} have employed pre-Sputnik images to identify transients, yielding intriguing results. In the present work, we focused on the Earth shadow filter, evaluating the effectiveness of the methods for detecting fast transients and flash trains using existing data from the Zwicky Transient Facility (ZTF).

Our first analysis focused on Sample A of 678 images, encompassing a small subset ($N$ = 262) of candidate transients in Earth's shadow from a total of 11,029 one-off candidate transients with clear PSFs observed in single ZTF images (shared by Igor Andreoni).  These candidates were constrained to durations shorter than 30 seconds, as given by the time difference between the previous or subsequent image captures (i.e., taken immediately before or after the given exposure). Among these cases, manual follow-up inspection has revealed at best several fast transients. Of special note is an uncatalogued object or set of objects shown in Figure \ref{weirdobject} and described in Table \ref{UncataloguedAsteroid}. Although the objects are located within a few degrees of the ecliptic, they do not appear in JPL's list of small bodies in the Solar system or the Minor Planet Center catalog as of April 2025. Moreover, if either pair of three detections represent an asteroid, then it moves across the sky at about 4 arcseconds per minute, which is several times faster than a main belt asteroid at opposition.

This leaves two possible interpretations. First, this case may be similar to the multiple clustered transient cases of unknown origin presented in \citep{Villarroel2021,Solano2024}.  A second possibility is that these detections represent a single object in motion (e.g., a remote-controlled human spacecraft or asteroid). This is a good illustration of why triangulation is necessary to disambiguate events of this kind.

We estimated the detection rate of one-off candidate transients. According to the IPAC server, a total exposure time of 6,415,320 seconds (approximately 1800 hours) resulted in the identification of 11,029 candidate transients. Of this, only 2.4\%---about 43 hours---was spent in the Earth’s shadow, during which the observations covered 67,492 square degrees (including overlapping fields). During this interval, we detected one candidate case for multiple transients in Figure \ref{weirdobject}, corresponding to a rate of $\sim 3 \times 10^{-7}$ cases per hour per square degree, or $\sim 0.01$ cases per hour across the whole sky. This estimate is based on a single event and should be interpreted with caution due to small-number statistics.

This is a factor of $10^{6}$ less than the number of fast flashes coming from reflections of space debris that has been measured previously \citep{Nir2021b,Corbett2020}. The multiple transient cases in \cite{Villarroel2021} were estimated to occur at a rate of $\sim 0.07$ transients hour$^{-1}$ sky$^{-1}$ and $\sim 0.27$ transients hour$^{-1}$ sky$^{-1}$ in \-\cite{Villarroel2022b}.

This small detection rate is not unexpected if the multiple transient candidates are attributed to solar reflections from artificial objects in geosynchronous orbits around the Earth, as suggested in \cite{Villarroel2021} and \cite{Villarroel2022b}. Further, the independent automated transient survey (sets B-C) carried out in Section \ref{sec:independent} (see total detections listed in Table \ref{tab_subsets}) suggests that there might be many more transients than were covered by the candidate transient alerts described in Section \ref{sec:setA}.

Nevertheless, severe methodological limitations affect the detection rate. Multiple transients were observed in images with 50 minutes of exposure in previous studies \citep{Villarroel2021,Solano2024}, whereas ZTF images typically only able to capture one-off transients, often in sets containing a few consecutive 30-second exposures of the same field.  Unfortunately, the time spent on exposure per field is typically only 2--3 minutes before the telescope moves to a very different part of the sky, which makes it difficult to capture a process that could be spread out over an hour or more. Therefore, the likelihood of observing multiple transients with the ZTF is not high.  The previous findings of multiple transients in the Palomar survey used approximately 50 minutes -- 1 hour of exposure \citep{Villarroel2021,Solano2024}. To obtain a reasonable chance of detection in ZTF, but also of distinguishing between the effects of ``multiple transients'' versus the movement of an asteroid, an equally long time window would need to be necessary,  resulting in hundreds of consecutive exposures of 30 seconds each---a condition not met in this study.

The automated survey of Samples B--C using {\it NEOrion} made possible the detection and study of many other interesting objects. We detected thousands of point sources of interest (as potential flashes), but also many streaks, which could form the focus of a follow-up study. The streaks shown in Figure \ref{coolstreaks} are within several degrees of the shadow center (Table \ref{filenames}). Streaks that cross the entire image, such as the one in the upper left of Figure \ref{coolstreaks}, are probably meteors. The example shown in the upper right is contained entirely within a single image frame.  The most likely explanation is a meteor striking the atmosphere at a steep angle.  The FWHM of the cross-section is consistent with stellar psfs in the scene, suggesting an envelope of ionized gas measuring less than 0.5 m in diameter if located in the upper atmosphere. Alternatively, this could be a luminous object in Earth's shadow orbiting at an elevation of approximately 20,000 km. Without an estimate of the parallax, we cannot say for sure whether the estimated distance based on the angular speeds matches expectations consistent with the more prosaic explanation. With the help of instantaneous spectra as described by e.g. \cite{Marcy2022a}, we could determine whether the luminosity derives from reflection or emission (and if so, what kind).

The methodological limitations of this study will be mitigated by the new ExoProbe project, see \cite{ExoProbe2023}. The project aims to build a network of telescopes with high-resolution Complementary Metal Oxide Semiconductor (CMOS) cameras to search for ET artifacts and probes in the inner Solar System, in search of short flashes (subsecond -- second) associated with technological objects of potential extraterrestrial origin. A series of short exposure images (1s) will be taken at multiple telescopes simultaneously during an hour at a time, searching for any flash that appears in several telescopes, so that we instantly obtain a parallax (and hence the location) of such short transients. The parallax information is absolutely essential in order to estimate true distance. This will tell us whether objects like those in Figures \ref{weirdobject}, \ref{coolstreaks} and \ref{Ecliptic} are at the distances indicated by the angular speeds. The ExoProbe project enables us to search for, locate, and instantly validate and verify the results of such detections.

Searches within the Earth's shadow offer an exciting opportunity as they filter out solar reflections from most human-made objects. In the ExoProbe project, we will also search for elongated streaks in the shadow. 
With the exceptions discussed earlier, human satellites are not expected to emit any intrinsic light in the optical band. Therefore, one of the most effective methods for identifying an ET artifact or probe is by thoroughly exploring the shadow in searches for pulsing and continuous light. Long-term follow-up study of any such objects may be necessary to rule out prosaic explanations like those earlier mentioned.  The significance of any discovery will be further enhanced if any of the objects exhibit unusual spectra or unusual motion.

Studies that make use of instruments with significantly higher time resolution and a ranging capability, such as the new ExoProbe instruments that will search in Earth's shadow, are necessary to find and convincingly identify technological objects of ET origin in near-Earth space. 

\section{Data availability}

The ZTF image data underlying this article are publicly available from the IPAC Infrared Science Archive at \url{https://irsa.ipac.caltech.edu}.

\section{Acknowledgments}

B.V. wishes to thank Geoff Marcy, Igor Andreoni and Guy Nir for kind and helpful discussions and suggestions doing the course of this work. The paper has further profited from helpful discussions with Peter McCullough, Kevin Krisciunas, Axel Brandenburg and Rudolf Bär.

ExoProbe is supported by an anonymous donor, to whom we are deeply grateful.

B.V. is also funded by the Swedish Research Council (Vetenskapsr\aa det, grant no. 2024-04708) and supported by the The L’Or\'{e}al - UNESCO For Women in Science International Rising Talents prize.

This work makes use of observations from the Zwicky Transient Facility project. ZTF is supported by the National Science Foundation under Grant No. AST-2034437 and a collaboration including Caltech, IPAC, the Weizmann Institute for Science, the Oskar Klein Center at Stockholm University, the University of Maryland, the University of Washington, Deutsches Elektronen-Synchrotron, the University of Wisconsin, Trinity College Dublin, Lawrence Livermore National Laboratories, IN2P3, the University of Warwick, Ruhr University Bochum, and Northwestern University.

\bibliographystyle{mnras}
\bibliography{ss-seti_bv.bib}

%%%%%%%%%%%%%%%%%%%%%%%%%%%%%%%%%%%%%%%%%%%%%%%%%%

%%%%%%%%%%%%%%%%% APPENDICES %%%%%%%%%%%%%%%%%%%%%

%%%%%%%%%%%%%%%%%%%%%%%%%%%%%%%%%%%%%%%%%%%%%%%%%%

% Don't change these lines
\bsp	% typesetting comment
\label{lastpage}
\end{document}